\documentclass[twocolumn,prb,amsmath,showpacs,amssymb,citeautoscript,floatfix]{revtex4}
\def\etal{{ \it et al. }}
\def\prb{{Phys. Rev. B }}

\usepackage{epsfig}
\usepackage{subfig}
\usepackage{graphicx} 
\begin{document}
\title{Half-metallicity and magnetism of GeTe doped with transition metals V, Cr and Mn: 
a theoretical study from the viewpoint of application in spintronics 
}
\author{Y. Liu}
\affiliation{College of Science, Yanshan University, Qinhuangdao,
Hebei 066004, China; Physics Department, Brock University, St.
Catharines, Ontario L2S 3A1, CANADA}
\author{S.K. Bose}
\email[author to whom correspondence should be sent:] {sbose@brocku.ca}
\affiliation{Physics Department, Brock University, St. Catharines,
Ontario L2S 3A1, CANADA }
\author{J. Kudrnovsk\'{y}}
\affiliation{Institute of Physics, Academy of the Sciences of the Czech Republic, Na Slovance 2,
182 21 Prague 8, Czech Republic}

\begin{abstract}
This work presents results for the magnetic properties of the compound GeTe doped with $3d$ transition metals V, Cr and Mn
from the viewpoint of potential application in spintronics. We report a  systematic
density-functional study of the electronic structure, magnetic and cohesive properties of these ternary 
compounds in both rock salt and zinc blende structures. In both cases, it is the Ge sublattice that is doped with the three
transition metals. Some of these compounds
are found to be  half-metallic at their optimized cell volumes. For these particular cases, 
we calculate both exchange interactions and the Curie
temperatures in order to provide some theoretical guidance to experimentalists trying to fabricate materials suitable for spintronic devices.
Discussions relating our results to the existing experimental studies are provided whenever applicable and appropriate.
Apparent discrepancy between experimental observations and our theoretical result for the case of Mn-doping is discussed in detail, pointing out 
various physical reasons and possible resolutions of the apparent discrepancy.
\end{abstract}
\maketitle

\section{Introduction}
In recent years GeTe thin films doped with $3d$ transition metals
have received considerable attention from experimentalists
interested in the field of magnetic
semiconductors\cite{Fukuma2001,Fukuma2003-1,Fukuma2003-2,Fukuma2006-1,Fukuma2006-2,Fukuma2006-3,Fukuma2008,Chen2008}.
Ferromagnetic order was observed for the Cr, Mn, and Fe doped films,
whereas Ti, V, Co and Ni doped films were found to be
paramagnetic\cite{Fukuma2003-1}. The Curie temperatures $T_c$ of
these thin films have been found to depend on the type and
concentration of the dopants, with a high value of 140 K, reported
for Ge$_{1-x}$Mn$_x$Te for $x$=0.51\cite{Fukuma2001}. Later on this
group of researchers  reported even higher value of  $T_c\sim$ 190 K
in dilute magnetic semiconductor (DMS)
Ge$_{0.92}$Mn$_{0.08}$Te\cite{Fukuma2008}. $T_c$s around 200-250 K have been
reported by Fukuma \etal  \cite{Fukuma2006-2} for Ge$_{1-x}$Cr$_x$Te thin
films for low values of $x$ ($\leq 0.1$). The thin films in all
these studies were of predominantly rock salt structure. For
spintronic applications, the desired properties of such ferromagnets
are half-metallicity and relatively high Curie temperatures.
Possibility  of half-metallicity in Cr- and V-substituted GeTe bulk
compounds has been reported by Zhao \etal\cite{Zhao2006}, based on
theoretical calculations using linear augmented plane waves (LAPW)
WIEN2k code for some ordered Ge-V-Te and Ge-Cr-Te compounds. These
authors left the issue of the Curie temperature unattended, as only
the electronic band structure, density of states and charge density
were studied, and not the exchange interaction in these compounds. A
theoretical study of the electronic structure of Ge$_{1-x}$Mn$_x$Te,
based on the generalized gradient approximation plus Hubbard U
(GGA+U), has been presented by Ciucivara\etal\cite{Ciucivara2007}
Among the magnetic properties only the magnetic moment was
considered, the issue of the exchange interaction and $T_c$ was not
addressed. Recently a large number of experimental studies related to magnetism and
magnetic transitions in several Ge-Te based alloys, such as Ge-Cr-Te, Ge-Mn-Te and Ge-Fe-Te) have
been reported.
\cite{Fukuma2007,Lim2009,Lechner2010,Tong2011,Fukuma2008b,Zvereva2010,Lim2011a,Lim2011b,Lim2011c,Hassan2011,Kilanski2010}
In addition, there is substantial amount work on DMSs in
general (for a review see [\onlinecite{Sato2}]), where the
magnetic properties arise via percolation effects among a very small
concentration of the magnetic atoms. For a discussion on magnetic percolation in DMSs
readers may consult the work by Bergqvist \etal \cite{Bergqvist}

In view of the large body of existing experimental work  as
mentioned above, we have embarked on a systematic theoretical study
of the GeTe system doped with $3d$ transition metal atoms. The most
commonly occurring structure for both bulk and thin films of these
compounds is the NaCl or the rock-salt (RS) structure. However, from the
theoretical viewpoint an equally interesting structure to study is
the zinc blende (ZB) structure. Both RS and ZB are fcc-based, but
differ in terms of the distance between the Ge and Te atoms. Thus
irrespective of whether one dopes the Te- or the Ge-sublattice with
a magnetic ($3d$) atom, the magnetic effects are expected to differ
because of different levels of hybridization between the $3d$
orbitals and the $s$ and $p$ orbitals of the Ge or Te atoms.
Experimentally it may be possible to grow both RS and ZB structures,
even though the ground state structure appears to be RS. Thus we
have studied the electronic and magnetic properties of both these
structures for different concentrations of the magnetic V, Cr, and
Mn atoms. We have employed supercell method as well as the coherent
potential approximation (CPA) to study the effect of doping at
various concentrations. The supercell calculations are carried out
using the full potential linear augmented plane wave (FP-LAPW)
method using the WIEN2k code\cite{Blaha-wien2k}. The CPA calculations are carried out
within the frame-work of the linear muffin-tin orbital (LMTO) method
using the atomic sphere approximation
(ASA)\cite{Kudrnovsky1990,Turek97}. In a limited number of cases,
when the half-metallicity is predicted on the basis of both
supercell and CPA calculations, we have provided results for
exchange interactions and Curie temperatures for these alloys.

\section{Computational details}
The crystal structures of the ternary GeTe based compounds
Ge$_{1-x}$TM$_x$Te (TM stands for transition metal atoms V, Cr and Mn) were constructed from the unit cell of RS and 
ZB structures as follows.  The doping levels of 
x=0.25 or 0.75 were achieved by replacing the Ge atoms at the vertex site or
face-center sites, respectively, of the RS/ZB unit cell , with the TM atoms.
 Both cases (x=0.25 or 0.75) have the same space group ($Pm\bar{3}m$, or 221) for the RS structure,
and ($P\bar{4}3m$, or 215) for ZB. 
The x=0.5 doping  is obtained by
replacing the Ge atoms at the four compatible face-center sites.
 In this way we generate  tetragonal structure ($P\bar{4}m2$, 115)
for the ZB case and ($P4/mmm$, 123) for RS. This procedure of realizing the
doping levels results in minimal primitive cells, with the highest possible symmetry. 
Numerous semiconductors are known to crystallize in  the ZB or RS
structures. As such these ternary compounds should be compatible with a large number of
semiconductors.

Calculations were carried out within the framework of the density-functional theory (DFT)\cite{Kohn-pr-136-B864-1964},
 using the  WIEN$2$k\cite{Blaha-wien2k} code based
on the full-potential linear augmented plane wave plus local orbitals
method. The
generalized gradient approximation (GGA) proposed by Perdew \etal
was used for exchange and correlation potentials\cite{Perdew-prb-45-13244-1992,perdew-prl-77-3865-1996}. We consider full
relativistic effects for the core states and use the scalar-relativistic
approximation for the valence states, neglecting the spin-orbit
coupling. The latter is known to produce only a small effect on the results that are of interest in this work,
e.g., density of states and the energy gaps. 

As a note to practitioners of this code, we used
$3000$ K points (Monkhorst-Park grid\cite{Monkhorst-prb-13-5188-1976}) 
 for the Brillouin-zone integration, set
$Rmt*Kmax=8.0$ and carried out the angular momentum expansion up to $l_{max}=10$ in
the muffin tins, and used $G_{max}=14$ for the charge density.
 All core and valence
states are treated self-consistently. Convergence with respect to basis set and
k-point sampling was carefully verified. 
The self-consistent calculation is allowed to stop only when the integrated
charge difference per formula unit, $\int|\rho_{n}-\rho_{n-1}|dr$, 
for the input and output charge densities $\rho_{n-1}$ and  $\rho_{n}$ is
less than $10^{-4}$. In the calculation of Ge-X-Te in different structures, 
the muffin-tin (MT) radii are chosen to be $2.3$, $2.2$ and $2.5$ a.u. for Ge, X (X=V, Cr, Mn)
 and Te atoms, respectively.

\section{Electronic structure}

\subsection{Results of supercell calculations using the FP-LAPW method}

For each case equilibrium lattice parameter was obtained by minimizing the total
energy with respect to the cell volume. All electronic properties such as 
the density of states (DOS), energy bands and magnetic moments were then
calculated for the equilibrium lattice parameters. Among all the
ternary TM compounds with the doping levels considered, we find 9 half-metallic (HM)
cases: one Mn-doped, three Cr-doped, and three V-doped cases for the RS structure; and one Cr-doped
and one V-doped  cases for the ZB structure (see Tables I-IV).
There is a small drop in the value of the equilibrium lattice parameter with increasing dopant
concentration in all cases. This is supported by the measurements of lattice parameters for the thin films
of Ge$_{1-x}$Mn$_x$Te and Ge$_{1-x}$Cr$_x$Te grown in RS structure \cite{Fukuma2006-2,Fukuma2001}. 
 Bulk moduli, calculated by using Birch-Murnaghan equation of state \cite{Murnaghan-pnas-30-5390-1944},
 are found to increase with the doping level for V- and Cr-doping.
For Mn-doping this change is non-monotonic.
The equilibrium lattice constants,
X-Te bond lengths, and bulk moduli, magnetic moments, minority-spin gaps,
 and half-metallic gaps for the ternary compounds in RS structure are summarized in Tables I and II.
The same quantities for the ternary compounds in ZB structure are summarized in Tables III and IV.
In general, the band gap tends to increase with decreasing lattice parameter. This is understandable as the increased hybridization
due to decreasing inter-atomic distances leads to increased separation between the energies of the bonding and anti-bonding states.
\begin{table}
\caption{\label{I} Results obtained via the WIEN2k code for the RS compounds Ge$_{4-n}$X$_{n}$Te$_4$ (X=V,Cr,Mn) : 
the equilibrium lattice constants $a$, the
X-Te bond length L$_{XTe}$, bulk modulus $B$, minority-spin gaps G$_{MIS}$ (eV), and half-metallic gaps
G$_{HM}$ (eV). All results shown are obtained using GGA (see text).  }

\begin{ruledtabular}
\begin{tabular}{cccccccccccccc}
compounds  &$a$(\AA)  &  $L$(\AA) &$B$(GPa)  & $G_{MIS}$(eV) &$G_{HM}$(eV)   \\
\hline
Ge$_3$V$_{1}$Te$_4$  &5.9283 &2.9642  &51.3   &0.599  &0.097 \\
Ge$_{2}$V$_{2}$Te$_4$  &5.8497 &2.9249  &55.8   &0.816  &0.170 \\
Ge$_{1}$V$_{3}$Te$_4$  &5.7865 &2.8933  &59.7   &1.007  &0.213 \\
Ge$_{3}$Cr$_{1}$Te$_4$ &5.9325 &2.9662  &51.4   &0.571  &0.220 \\
Ge$_{2}$Cr$_{2}$Te$_4$ &5.8579 &2.9290  &54.9   &0.789  &0.382 \\
Ge$_{1}$Cr$_{3}$Te$_4$ &5.7860 &2.8930  &58.2   &0.980  &0.371 \\
Ge$_{3}$Mn$_{1}$Te$_4$ &5.9638 &2.9819  &49.7   &0.710  &0.200 \\
Ge$_{2}$Mn$_{2}$Te$_4$ &5.9139 &2.9568  &46.1  &-  &- \\
Ge$_{1}$Mn$_{3}$Te$_4$ &5.8413 &2.9206  &43.1   &-  &- \\
\end{tabular}
\end{ruledtabular}
\end{table}

\begin{table}
\caption{\label{II} Magnetic moment per magnetic atom (f.u.) in units of
bohr magneton  $\mu_B$,
for the RS compounds
 Ge$_{4-n}$X$_{n}$Te$_4$ (X=V, Cr, Mn), at their respective equilibrium
volumes, obtained by using the FP-LAPW method (WIEN2k). Moments associated with
X, Te, Ge muffin-tin spheres and the interstitial region (Int) are shown separately.}

\begin{ruledtabular}
\begin{tabular}{cccccccccccccccc}
compounds                 & X($\mu_B$)  & Te($\mu_B$)    & Ge($\mu_B$)  & Int($\mu_B$)   & Tot ($\mu_B$)   \\
\hline
Ge$_3$V$_{1}$Te$_4$  &2.463 &-0.050 &0.026 &0.624  &3.000  \\
Ge$_{2}$V$_{2}$Te$_4$ &2.482 &-0.053 &0.043 &0.606  &3.000  \\
 Ge$_{1}$V$_{3}$Te$_4$ &2.497 &-0.076 &0.048 &0.589  &3.000  \\
Ge$_{3}$Cr$_{1}$Te$_4$ &3.580 &-0.065 &0.017 &0.576  &4.000  \\
Ge$_{2}$Cr$_{2}$Te$_4$ &3.575 &-0.065 &0.031 &0.565  &4.000  \\
Ge$_{1}$Cr$_{3}$Te$_4$ &3.567 &-0.120 &0.047 &0.558  &4.000  \\
Ge$_{3}$Mn$_{1}$Te$_4$ &4.243 &0.035 &0.024 &0.583  &5.000  \\
Ge$_{2}$Mn$_{2}$Te$_4$ &4.226 &0.032 &0.053 &0.561  &4.947  \\
Ge$_{1}$Mn$_{3}$Te$_4$ &4.140 &0.050 &0.083 &0.494  &4.167  \\
\end{tabular}
\end{ruledtabular}
\end{table}

\begin{table}
\caption{\label{III} 
Results obtained via the WIEN2k code for the ZB compounds Ge$_{4-n}$X$_{n}$Te$_4$ (X=V,Cr,Mn) : the equilibrium lattice constants $a$, the
X-Te bond length L$_{XTe}$, bulk modulus $B$, minority-spin gaps G$_{MIS}$ (eV), and half-metallic gaps
G$_{HM}$ (eV). All results shown are obtained using GGA (see text).}

\begin{ruledtabular}
\begin{tabular}{cccccccccccccc}
compounds  &a(\AA)  &  L(\AA) &B(GPa) & G$_{MIS}$(eV) &G$_{HM}$(eV)   \\
\hline
Ge$_3$V$_{1}$Te$_4$  &6.5842  &2.8510  &27.7  &-  &-  \\
Ge$_{2}$V$_{2}$Te$_4$  &6.4316  &2.7953  &33.8  &-  &- \\
Ge$_{1}$V$_{3}$Te$_4$  &6.3109  &2.7327  &42.2  &1.061 &0.482\\
Ge$_{3}$Cr$_{1}$Te$_4$ &6.5950  &2.8556  &30.4  &-  &-  \\
Ge$_{2}$Cr$_{2}$Te$_4$ &6.4347  &2.7866  &34.0  &-  &- \\
Ge$_{1}$Cr$_{3}$Te$_4$ &6.3280  &2.7401  &38.8  &1.144 &0.253\\
Ge$_{3}$Mn$_{1}$Te$_4$ &6.6405  &2.8754  &29.1  &-  &-  \\
Ge$_{2}$Mn$_{2}$Te$_4$ &6.5305  &2.8281  &30.1  &-  &- \\
Ge$_{1}$Mn$_{3}$Te$_4$ &6.4222  &2.7809  &26.5  &- &-\\
\end{tabular}
\end{ruledtabular}
\end{table}

\begin{table}
\caption{\label{IV} 
Magnetic moment per magnetic atom (f.u.) in units of
bohr magneton  $\mu_B$,
for the ZB compounds
Ge$_{4-n}$X$_{n}$Te$_4$ (X=V, Cr, Mn), at their respective equilibrium
volumes, obtained by using the FP-LAPW method (WIEN2k). Moments associated with
X, Te, Ge muffin-tin spheres and the interstitial region (Int) are shown separately.}

\begin{ruledtabular}
\begin{tabular}{cccccccccccccccc}
compounds                 & X ($\mu_B$)  & Te($\mu_B$)    & Ge($\mu_B$)  & Int($\mu_B$)   & Tot($\mu_B$)    \\
\hline
 Ge$_3$V$_{1}$Te$_4$ &2.535  &-0.079  &0.022  & 0.719  &3.007 \\
 Ge$_{2}$V$_{2}$Te$_4$ &2.510  &-0.109  &0.027  & 0.690  &3.009 \\
 Ge$_{1}$V$_{3}$Te$_4$ &2.463  &-0.113  &0.040  & 0.674  &3.000 \\
Ge$_{3}$Cr$_{1}$Te$_4$ &3.706  &-0.079  &0.013  & 0.675  &4.103 \\
Ge$_{2}$Cr$_{2}$Te$_4$ &3.625  &-0.126  &0.014  & 0.658  &4.045 \\
Ge$_{1}$Cr$_{3}$Te$_4$ &3.560  &-0.145  &0.021  & 0.627  &4.000 \\
Ge$_{3}$Mn$_{1}$Te$_4$ &4.171  &0.024  &0.033  & 0.634  &5.003 \\
Ge$_{2}$Mn$_{2}$Te$_4$ &4.190  &0.054  &0.068  & 0.635 &5.002\\
Ge$_{1}$Mn$_{3}$Te$_4$ &4.191  &0.056  &0.078  & 0.597  &4.889 \\

\end{tabular}
\end{ruledtabular}
\end{table}

 Tables \ref{II} and \ref{IV} show that most of the moments are associated with the muffin-tins around the 
TM atoms and the interstitial space. The orbitals associated with Ge and Te atoms have
little spin polarization. While the values of the TM and interstitial moments are dependent on the muffin-tin radii, 
the integer values of the total moments shown in the last column
clearly identify the half-metallic cases.

Note that in all cases studied and reported in Tables \ref{I}-\ref{IV}, the RS structure has 
lower energy than the ZB, indicating the equilibrium bulk phase to be RS
at low temperatures. However, it might be possible to grow  thin films of these 
compounds in ZB structure under suitable conditions.  Therefore, there is some merit
in comparing the magnetic properties of these two fcc-based phases. So far all thin 
films of these compounds seem to have been grown in RS structure or with small 
rhombohedral distortions from this structure\cite{Fukuma2001,Fukuma2003-1,Fukuma2008,Fukuma2008b}.

\begin{figure*}
\center \includegraphics[width=14cm]{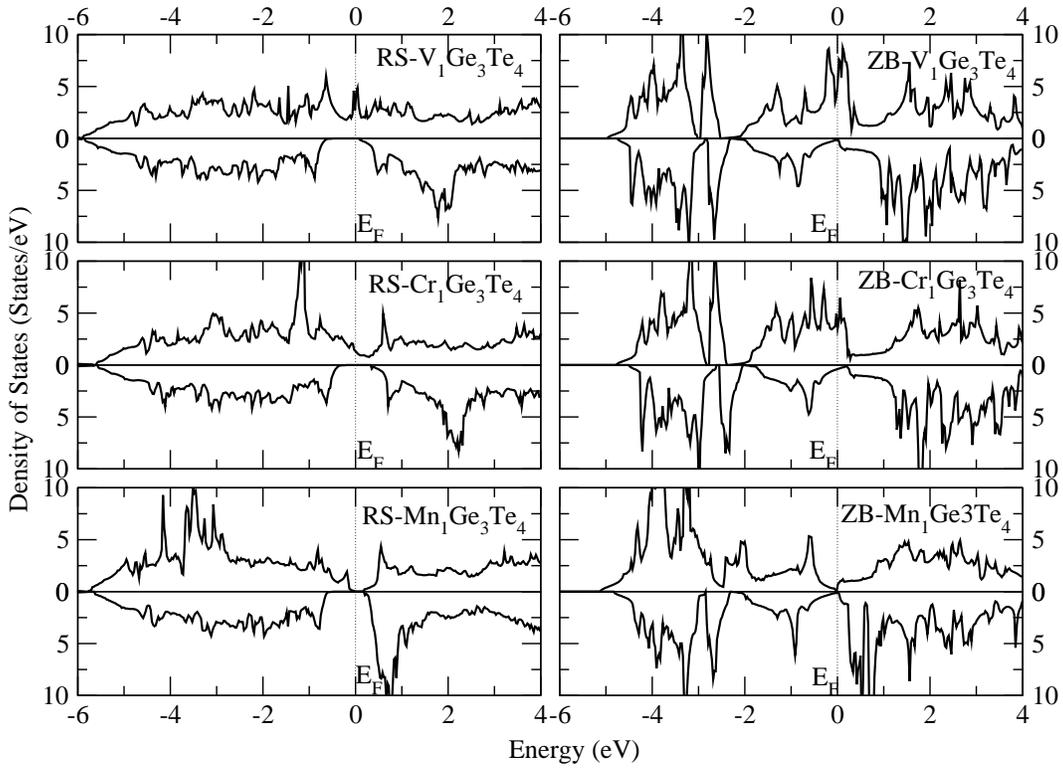}
\caption {Densities of states in Ge$_3$TM$_{1}$Te$_4$ (TM=V,Cr,Mn) alloys in RS and ZB structures.
}
\label{f1}
\end{figure*}

\begin{figure}
\center \includegraphics[angle=270,width=9cm]{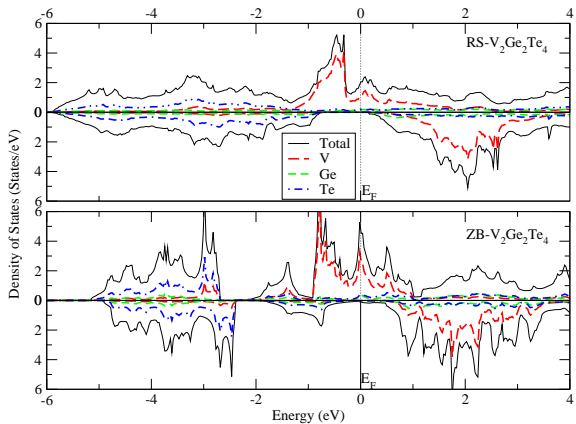}                  
\caption {(Color online) Total and component-resolved densities of states in V$_2$Ge$_2$Te$_4$  alloys in RS and ZB structures.
}
\label{f2}
\end{figure}

The densities of states for some these ordered compounds at their equilibrium lattice 
parameters are shown in Figs.(\ref{f1}-\ref{f2}).
These results clearly show  in which cases half-metallicity is most robust, 
i.e. the Fermi level is most widely separated from the band edges.
The partial atom-projected densities of states shown for Ge$_2$V$_2$Te$_4$, 
shows that the states at the Fermi level have a high TM character.
The situation should be similar for TM=Cr, Mn as well, with the amount of the TM character changing with TM concentration.
The 25\% Mn-doping case needs special attention, since the metallic character is either marginal (ZB structure) or absent (RS structure). 
For Ge-Te in the ZB structure, 25\% Mn-doping of the Ge-sublattice produces a borderline, i.e., zero gap, semiconductor or semimetal. For the
RS-structure, the same Mn-doping level of the Ge-sublattice produces a narrow gap semiconductor. As we will see in the next section, the nature of 
magnetism in these compounds is dictated by the availability of free carriers or holes. Thus the Mn-doping case will be considered in some 
additional detail.
 
The results presented in Tables I-IV and Figs. 1-2  are for ordered alloys, while the TM-doped thin films of GeTe are partially random,
in the sense that the Ge-sublattice sites are occupied randomly by Ge and TM atoms. We have studied the electronic structure of these
partially random alloys using the tight-binding linear muffin-tin orbital(TB-LMTO) method in conjunction with the CPA.\cite{Kudrnovsky1990,Turek97}
These calculations employed exchange-correlation potential of Vosko, Wilk and Nusair\cite{VWN}, an $s,p,d,f$ basis set, relativistic treatment of core
electrons and scalar-relativistic treatment of valence electrons. The results, particularly with respect to half-metallicity, were similar to those
of the ordered alloys, apart from expected smoothing of some peaks in the DOS. The gaps values were 
marginally lower and can be ascribed to the differences between local density approximation
(LDA) in TB-LMTO and GGA in FP-LAPW. The equilibrium lattice parameters were 2-4\% higher in TB-LMTO LDA calculations.

\section{Exchange interaction and Curie temperature}
\label{sec:exchange}
%\subsection{Mapping onto a Heisenberg Hamiltonian and related issues}
 \textit {First-principles} calculations of the thermodynamic properties of itinerant magnetic systems,
 via mapping \cite{Sandratskii,Pajda2001} of the zero temperature band energy onto a classical Heisenberg model:
\begin{equation}\label{e1}
H_\mathrm{eff} = - \sum_{i,j} \, J_{ij} \,
{\bf e}_{i} \cdot {\bf e}_{j} \ ,
\end{equation}
has been discussed in detail in our previous publications\cite{Bose2010,Liu2010}.
Here $i,j$ are site indices, ${\bf e}_{i}$ is the unit vector
pointing along the direction of the local magnetic moment at site
$i$, and $J_{ij}$ is the exchange interaction between the moments at
sites $i$ and $j$. The calculations are based on a mapping procedure due to
Liechtenstein \etal
\cite{Liechtenstein84I,Liechtenstein84II,Gubanov92}
The method was later extended to random  magnetic systems by Turek \etal, using CPA and the TB-LMTO method\cite{Turek2006}.
The exchange integral in Eq.(\ref{e1}) is given by
\begin{equation}\label{eq-Jij}
 J_{ij} = \frac{1}{4\pi} \lim_{\epsilon\rightarrow 0^+} Im \int tr_L \left[\Delta_i(z)g_{ij}^{\uparrow}(z)
\Delta_j(z)g_{ji}^{\downarrow}
\right] dz \; ,
\end{equation}
where $z=E+i\epsilon$ represents the complex energy variable, $L=(l,m)$, and $\Delta_i(z)= P_i^{\uparrow}
(z)-P_i^{\downarrow}(z)$, representing the difference in the potential
functions for the up and down spin electrons at site '$i$'. 
$g_{ij}^{\sigma}(z) (\sigma=\uparrow,\downarrow)$
represents the matrix elements of the Green's function of the medium for the up and down spin electrons.
For sublattices with disorder, this is a configurationally averaged Green's function, obtained via using the prescription of CPA.
It should be noted that the spin magnetic moments are included in the above definition of $J_{ij}$. Positive and negative values of 
$J_{ij}$ imply FM and AFM couplings, respectively, between the atoms at sites $i$ and $j$.

\begin{figure}
\center \includegraphics[width=8cm]{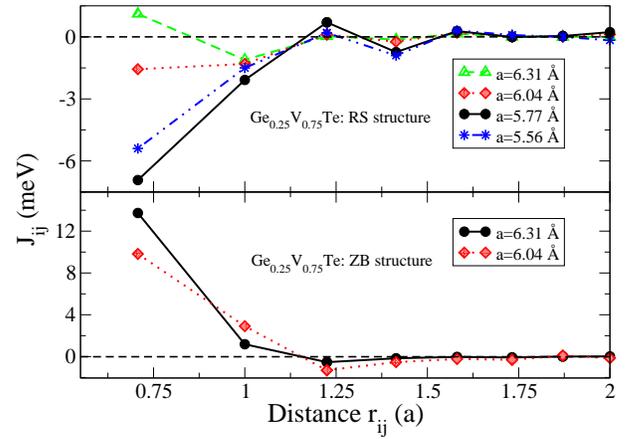}                 
\caption {(Color online) Exchange interaction between the V atoms in Ge$_{0.25}$V$_{0.75}$Te  as a function 
of interatomic distance expressed in units of lattice parameter $a$. Solid lines with
solid circles show the results for the equilibrium lattice parameters.
}
\label{f3}
\end{figure}

\begin{figure}
\center \includegraphics[width=8cm]{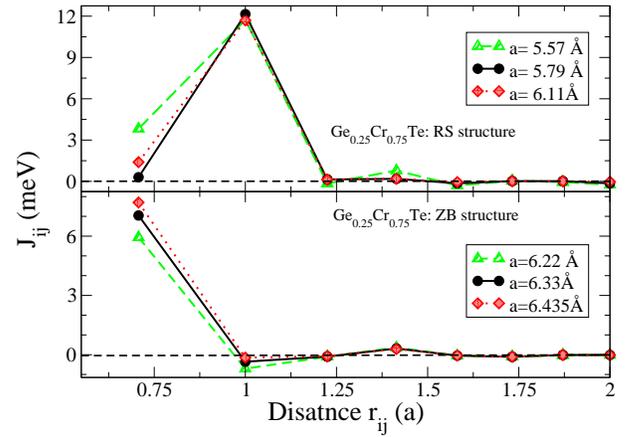}                 
\caption {(Color online) Exchange interaction between the Cr atoms in Ge$_{0.25}$Cr$_{0.75}$Te  as a function 
of interatomic distance expressed in units of lattice parameter $a$.
Solid lines with solid circles show the results for the equilibrium lattice parameters.}
\label{f4}
\end{figure}

\begin{figure}
\center \includegraphics[width=8cm]{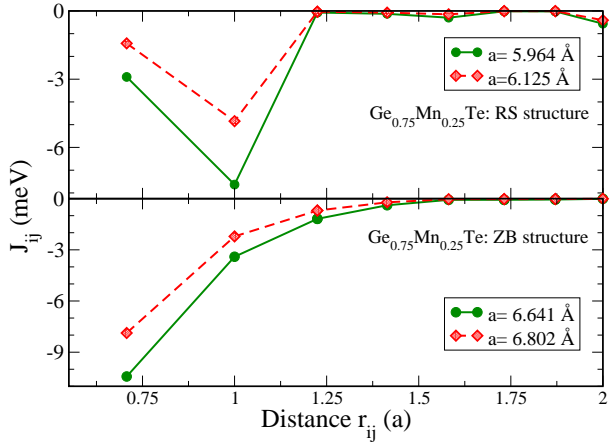}               
\caption {(Color online) Exchange interaction between the Mn atoms in Ge$_{0.75}$Mn$_{0.25}$Te  as a function 
of interatomic distance expressed in units of lattice parameter $a$.
Solid lines with
solid circles show the results for the equilibrium lattice parameters.}
\label{f4a}
\end{figure}

Existence of the HM gap in spin polarized electronic densities of states does not 
guarantee that the substance is actually ferromagnetic (FM).
The search for the ground magnetic state often needs to be guided by \textit{ab initio} 
calculation of the exchange interactions between various atoms.
In the above procedure outlined by Liechtenstein \etal \cite{Liechtenstein84I,Liechtenstein84II,Gubanov92}, 
exchange interactions are calculated by considering spin deviation
from a reference state. Negative exchange interactions resulting from  calculations based on a 
FM reference state would suggest instability of the assumed FM ground
state. With this mind, we have computed the exchange interactions for the most 
interesting cases, i.e. those promising robust HM states. 
The results for the exchange interactions are shown in Figs.(\ref{f3}-\ref{f4a})

In Figs. \ref{f3} and \ref{f4} we show results for the  cases with 75\% doping with Cr and V 
for the RS and ZB structures, as this level of doping yields largest values
of the minority spin and HM gaps. In Fig. \ref{f4a} we consider the 25\% Mn-doping case for 
the RS structure, the only Mn-doped case studied which shows the promise of
  half-metallicity. The solid lines (with circles) in these figures refer to the alloys with 
the equilibrium lattice parameters. In order to understand some trends,
we have also considered  lattice parameters above and/or below the equilibrium values. 
Ferromagnetic interactions are strongest in the Cr-doped ZB GeTe.
The Cr-doping case is most promising, as in both RS and ZB structures FM interactions dominate, 
and these remain FM with changes in the lattice parameter.
The next promising case is V-doping in the ZB structure, the RS counterpart showing 
strong antiferromagnetic (AFM) interactions, particularly at and around the
equilibrium lattice parameter. These AFM interactions weaken on both sides of the 
equilibrium lattice parameter value and become FM only at much higher volume.
 Fig.\ref{f4a} shows that all interactions up to many neighbor shells are AFM, and 
then die off to zero. Hence the ground state cannot be
FM. This conclusion needs to reinforced by examining the Lattice Fourier transformation 
of all the interactions. However, the preponderance of AFM interactions
would dictate that the ground state is perhaps AFM or of complex magnetic structure. 
This issue is further explored in the next section. For the 25\% Mn-doped solid the interactions are 
strongly AFM not only at the equilibrium lattice parameter, but also for
a substantial range of the lattice parameter around the equilibrium value. The interactions can 
become FM only at unrealistically large lattice parameter,
 while compression would result in stronger AFM interactions. While our theoretical calculations 
suggest that  25\% Mn-doping of the Ge-sublattice should lead 
to a ground state that is either AFM or of a more complex magnetic structure, some recent 
experimental results suggest ferromagnetism in these or similar Mn-doped
compounds. Thus we will discuss the Mn-doping case separately in order to shed some 
light on the apparent discrepancy between theoretical results and the available
experimental studies. In the following paragraphs we first present estimates of 
the Curie temperature for the Cr- and V-doped Ge-Te.


We have calculated the Curie temperature $T_c$  using both the mean-field approximation
(MFA) and the more accurate random-phase approximation
(RPA)\cite{Prange}. If the magnetic sublattice consists only of the magnetic atoms X, then in the MFA, the Curie
temperature is given by

\begin{equation}\label{e2}
k_{B} \, T_{c}^{\rm MFA} =
\frac{2}{3}\sum_{i \ne 0} \, J^{\rm X,X}_{0i} \, ,
\end{equation}
where the sum extends over all the neighboring shells and involves the exchange interactions
between the magnetic
atoms $X$.
MFA is known to grossly overestimate $T_c$.
A much more improved description of finite-temperature magnetism
is actually provided by the RPA. Again, if the magnetic sublattice consists only 
of the magnetic atoms X, then the RPA  $T_{c}$ given by
\begin{equation}\label{e3}
(k_{B} \, T_{c}^{\rm RPA})^{-1} = \frac{3}{2} \frac{1}{N} \;
\sum_{\bf q} \, [ J^{\rm X,X}({\bf 0})-J^{\rm X,X}({\bf q}) ]^{-1}
\, .
\end{equation}
Here $N$ denotes the order of the translational group applied and
$J^{\rm X,X} (\bf q)$ is the lattice Fourier transform of the
real-space exchange interactions $J^{\rm X,X}_{ij}$. In order to address the randomness in the Ge-TM sublattice, 
we have modified Eqs.(\ref{e2}) and (\ref{e3}) for our Ge$_{1-x}$X$_x$Te alloys
using virtual crystal approximation (VCA). This involves simply weighting the exchange integrals in Eq. (\ref{e1})
  by  $x^2$, where $x$ is the concentration of the X (TM) atoms. As a result, Tc's obtained 
from Eqs.(\ref{e2}) and (\ref{e3}) get multiplied by the
same factor. In this way the problem is formally reduced to a nonrandom case. This
approximation fails for low concentrations, below the percolation limit\cite{Bergqvist}. 
The error decreases monotonically for higher and higher concentrations.
In general, the VCA results may somewhat overestimate the Curie temperature.

\begin{figure}
\center \includegraphics[width=8cm]{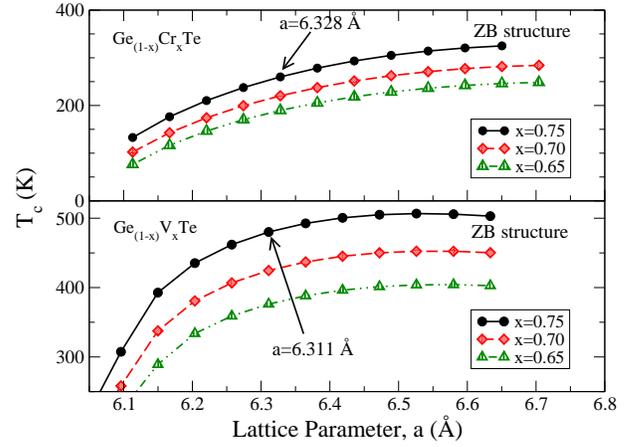}                  
\caption {(Color online) T$_c$ (via RPA together with VCA) versus lattice parameter $a$ in Ge$_{1-x}$X$_{x}$Te 
(X=V, Cr) alloys in ZB structure. The arrows indicate the values for the
equilibrium lattice parameters for $x$=0.75
}
\label{f5}
\end{figure}

\begin{figure}
\center \includegraphics[width=8cm]{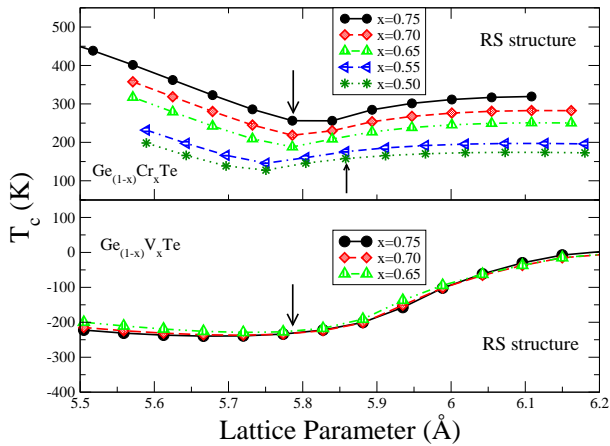}                  
\caption {(Color online) T$_c$ (via RPA together with VCA) versus lattice parameter $a$ in Ge$_{1-x}$X$_{x}$Te 
(X=V, Cr) alloys in RS structure. The arrows indicate the values for the equilibrium 
lattice parameters for $x$=0.75, and 0.50 for Ge$_{1-x}$Cr$_{x}$Te and $x$=0.75 for Ge$_{1-x}$V$_{x}$Te.
}
\label{f6}
\end{figure}

A problem arises in the computation of $T_c$ using either Eq.(\ref{e2}) or (\ref{e3}) when, in addition to the
robust moments on the magnetic atoms, there are induced moments on apparently non-magnetic atoms, interstitial spaces
or, in case of the LMTO method, empty spheres. This problem has been discussed in detail in our previous publication\cite{Bose2010}
and references cited therein.  
 As shown by Sandratskii \etal\cite{Sandratskii}
the calculation of $T_c$ using RPA is considerably more involved
even for the case where only one secondary induced interaction needs to be
considered, in addition to the principal interaction between the
strong moments. The complexity of the problem increases even for
MFA, if more than one secondary interaction is to be considered. 
In our case, the induced moments are small and as such interactions involving non-magnetic
atoms/spheres can be neglected in the first approximation. In addition, we are only interested
in a rough estimation of $T_c$, the object being able to determine which doping would lead to a higher 
 $T_c$ and hopefully be close to or above the room temperature. Our results, obtained by using
the RPA and considering only the TM moments in Eq.(\ref{e3}), and further modified in the spirit of VCA, are summarized Figs. \ref{f5} and \ref{f6}.
We show $T_c$ as a function of the lattice parameter for various relevant levels of doping. The equilibrium 
lattice parameter values are indicated with arrows.
Doping with Cr or V in the ZB structure for doping levels around 75\% is promising, with V-doping 
showing higher $T_c$ values than Cr-doping. V-doping in RS structure
is not recommended, while Cr-doping in RS structure leads to reasonably high values of $T_c$ for all 
levels of doping higher than 50\%. In a previous 
publication\cite{Liu2010} we have already presented a comparison of pure CrTe in ZB and RS structures, 
pointing out various advantages of considering RS CrTe over 
ZB CrTe from the viewpoint of spintronics applications.
\begin{figure*}
\centering
\subfloat[{\bf RS AFM[111]}]{\includegraphics[width=7.0 cm]{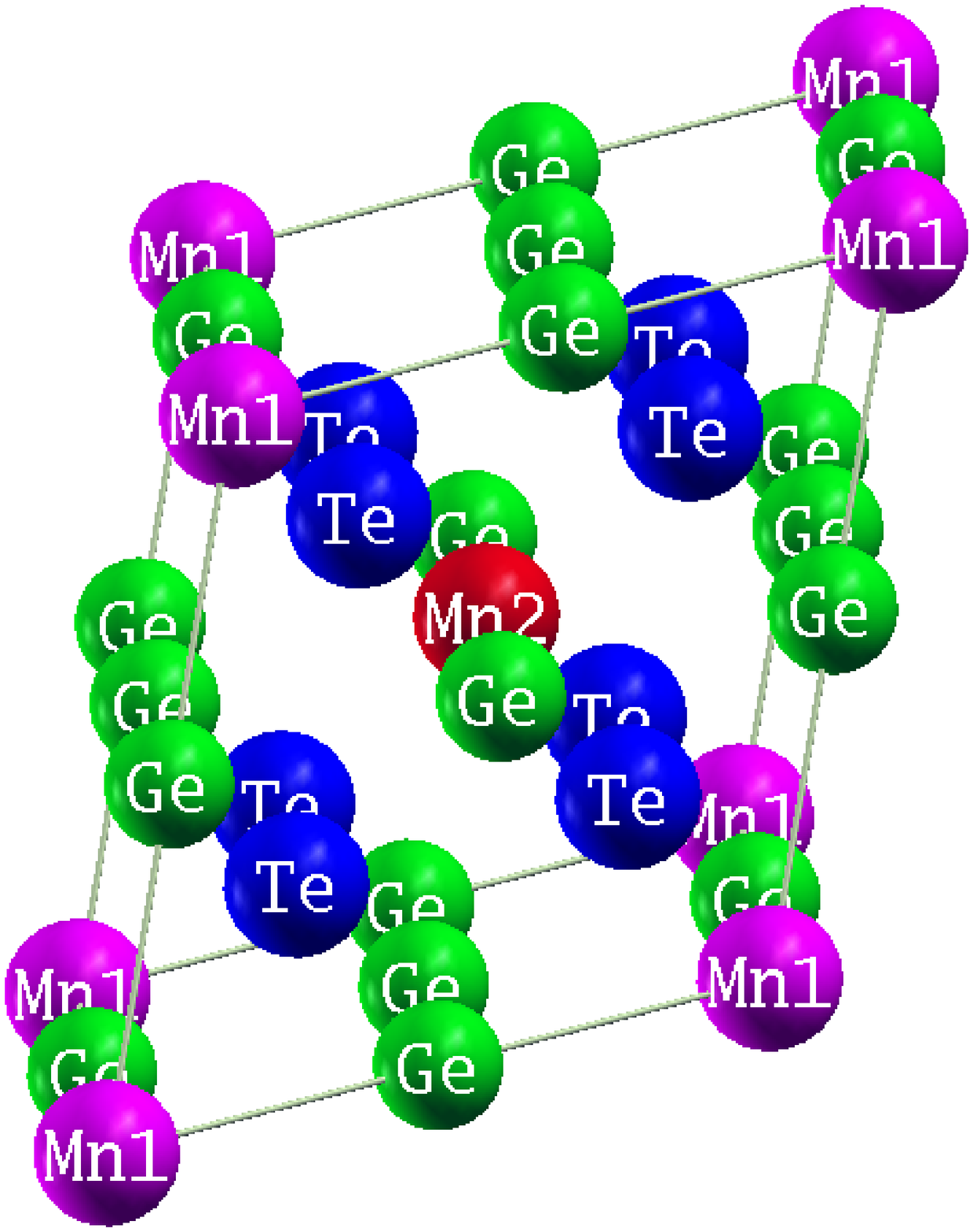}}     
\subfloat[{\bf ZB AFM[111]}]{\includegraphics[width=7.0 cm]{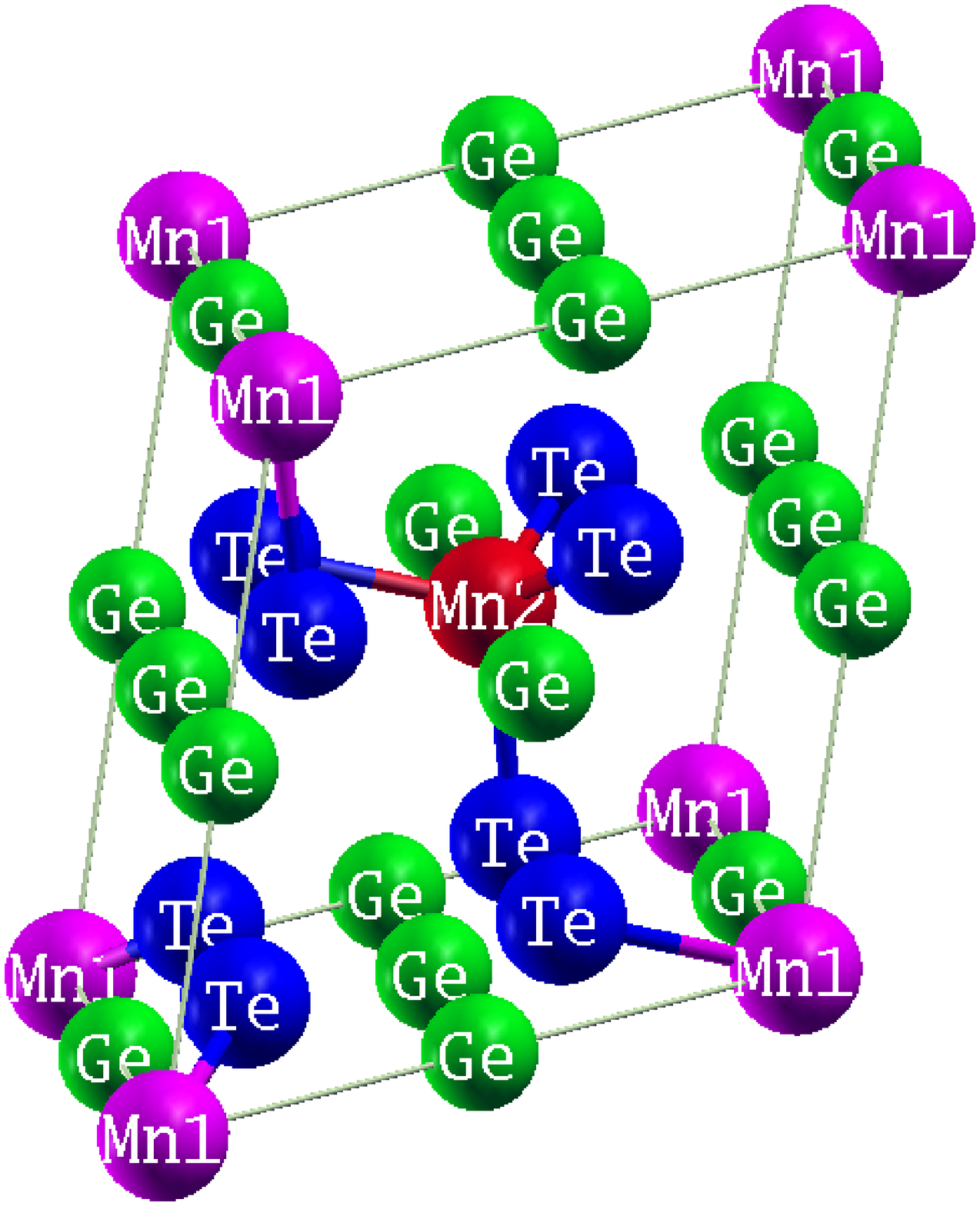}}     
\caption {(Color online) Unit cells of Ge$_3$Mn$_1$Te$_4$ in AFM [111] configuration.
}
\label{fig-AFM-RS_ZB}
\end{figure*}
\section{The case of $\rm{{\bf Ge}}$-$\rm{{\bf Mn}}$-$\rm{{\bf Te}}$}

Our results for the alloys Ge$_3$Mn$_1$Te$_4$ and Ge$_{0.75}$Mn$_{0.25}$Te need to be examined 
carefully in the context of the existing experimental studies, many of which
claim evidence of ferromagnetism over a  wide range of Mn concentration with $T_c$'s up to 100-200 K.  
Our LMTO-ASA-based exchange interactions 
indicate that the ground state should not be FM, at least for the 25\% Mn concentration case 
studied by us. These results are based on LDA. 
So first we have checked that the use of GGA and more accurate full-potential results are not 
drastically different. In Table \ref{V} we show the results for the FP-LAPW
(Wien2k) results obtained with GGA for the ordered Ge$_3$Mn$_1$Te$_4$ alloy. These results show that 
for both RS and ZB structures the AFM state 
energies are lower than the FM state energies, with the AFM [111] configuration having the lowest 
energy among the three AFM configurations considered. The AFM [111]
unit cells of Ge$_3$Mn$_1$Te$_4$, doubled with respect to the conventional FM unit cells, 
are shown in Fig. \ref{fig-AFM-RS_ZB}. 

\begin{figure}
\center \includegraphics[width=7cm]{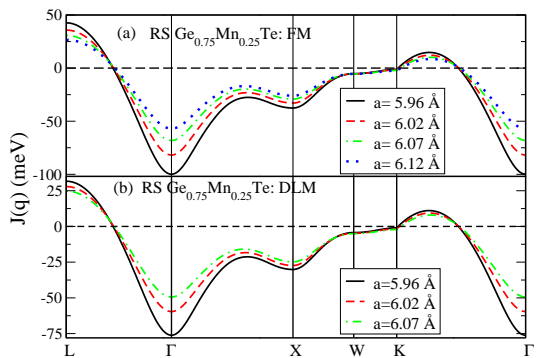}        
\caption {(Color online) Lattice Fourier transform of Mn-Mn interactions for various lattice parameters in RS 
Ge$_{0.75}$Mn$_{0.25}$Te for (a) FM and (b) DLM reference states. Solid lines represent
the results for the equilibrium lattice parameter.
}
\label{f7}
\end{figure}

In the RS structure the
energy difference of the AFM configurations, in particular the [111] case, is larger 
than for the ZB case. To examine the magnetic structure of the ground state we also
consider the lattice Fourier transform of the exchange interaction between the Mn atoms:
\begin{equation}
 J({\bf q})= \sum_{{\bf R}}J^{\rm Mn,Mn}_{0{\bf R}}\exp\left(i{\bf q}\cdot{\bf R}\right);
\end{equation}
where {\bf q} is a wave vector in the BZ of the fcc lattice. A maximum at the L point 
would point to the ground state being AFM [111], while that at the $\Gamma$ point
would indicate the ground state being FM. We have examined the ground state magnetic 
structure via $J({\bf q})$ using both FM and the disordered local moment (DLM)
\cite{Heine1,Heine2} reference states. Within the Stoner model, a nonmagnetic state above the Curie 
temperature $T_c$ is characterized by the vanishing of the local moments in magnitude. 
This obviously flawed description of the nonmagnetic state can be
improved by using the DLM model, 
where the local moments remain nonzero in magnitude above $T_c$, but disorder in  
their direction above $T_c$ causes global magnetic moment to vanish. For Mn this is known to 
be a good approximation. Within the collinear magnetic model, where all local axes of 
spin-quantization point in the same direction, DLM can be treated as a binary 
alloy problem\cite{Bose2010}.  The results for the RS structure, 
obtained by using the FM and DLM reference states are shown in sections (a) and (b) of Fig. \ref{f7},
respectively. Exchange interactions based on both FM and DLM reference states indicate that 
the ground state magnetic structure of RS Ge$_{0.75}$Mn$_{0.25}$Te is AFM [111].
\begin{table}
\caption{\label{V} 
Energy differences between ferromagnetic (FM) and various antiferromagnetic (AFM) configurations of
RS and ZB Ge$_3$Mn$_1$Te$_4$. The energies of the AFM unit cells are compared with 2 FM unit cells.
MO stands for magnetic order. The energy difference $\Delta$E of the AFM configurations with respect to the FM,
in the last column, are given per formula unit.}
\begin{ruledtabular}
\begin{tabular}{ccc}
MO                 & Energy (eV)  & $\Delta$E (meV)    \\
\hline
RS structure: & & \\

FM  & -942305.912 & 0 \\
 AFM001 & -1884611.922 & -49\\
 AFM110 &-1884611.981  &-79 \\
AFM111 & -1884612.033 & -105 \\
\hline
ZB Structure: & &\\

FM & -942304.475 & 0 \\
AFM001 & -1884608.979 &-14 \\
AFM110 & -1884609.006  &-28 \\
AFM111 & -1884609.018 & -33\\
\end{tabular}
\end{ruledtabular}
\end{table}

\begin{figure}
\center \includegraphics[width=7cm]{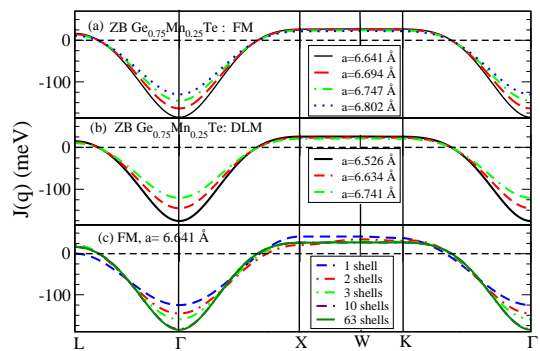} 
\caption {(Color online) Lattice Fourier transform of Mn-Mn interactions in ZB Ge$_{0.75}$Mn$_{0.25}$Te for 
(a) FM and (b) DLM reference states and  for various lattice parameters.
Section (c) shows variation with respect to the number of neighbor shells. Results for 10 
shells of neighbors are indistinguishable from those of 63 shells used in all our calculations.
Solid lines in (a) and (c) represent the results for the equilibrium lattice parameter.
}
\label{f9}
\end{figure}

However, the situation for ZB Ge$_{0.75}$Mn$_{0.25}$Te is not so clear. Both DLM and FM 
reference states yield $J({\bf q})$ curves with a broad maximum enclosing the symmetry
points X, W, and K. In addition, the values of $J({\bf q})$ at these points is marginally 
higher than that at the L point (sections (a) and (b) of Fig. \ref{f9}). What is certain is that
 the ground
state is not FM, with $J({\bf q})$ being a minimum at the $\Gamma$-point. Thus the ground state 
for ZB Ge$_{0.75}$Mn$_{0.25}$Te may involve a complex magnetic structure or
the substance may enter a spin-glass state at low temperatures. In order to ensure that our 
results do not suffer from a convergence problem, we have looked at the $J({\bf q})$
values as a function of increasing shells of neighbors. The results for ZB 
Ge$_{0.75}$Mn$_{0.25}$Te with FM reference state are shown in section (c) of Fig.\ref{f9}. Results for 10 shells of
neighbor are indistinguishable from those for 63 shells considered in all our calculations, 
indicating that our results are well-converged.

Experimentally, Ge$_{1-x}$Mn$_x$Te alloys, with X=Cr, Mn, Fe have been reported to exhibit ferromagnetic behavior
\cite{Fukuma2007,Lim2009,Lechner2010,Tong2011,Fukuma2008b,Zvereva2010,Lim2011a,Lim2011b,Lim2011c,Hassan2011,Kilanski2010}
 over wide range of the concentration $x$.
The structure of these compounds, grown as thin films, is usually found to be RS or small deviations 
from this. For Ge$_{1-x}$Mn$_x$Te the structure
 is found to be  rhombohedrally distorted NaCl for small Mn concentrations, with the distortion vanishing for $x > 0.18$\cite{Fukuma2001}. 
These studies also report presence of carriers in the samples, with transport measurements most often indicating p-type carriers
\cite{Fukuma2001,Fukuma2003-1,Fukuma2008,Fukuma2008b}, independent of temperature. Fukuma and co-workers identify the carriers as holes
\cite{Fukuma2001,Fukuma2008,Fukuma2008b}. Depending on carrier concentration, brought about 
by varying Mn concentration or hydrostatic pressure, 
T$_c$ may vary non-monotonically\cite{Fukuma2001,Lim2011b}. Highest reported T$_c$ is around 130-140 K 
for Mn concentration $x$ slightly over 50\% in the Ge-Mn
sublattice, and for $x=0.25$  T$_c$ is $\sim$ 70 K \cite{Fukuma2001}. Other authors report  T$_c$ values more or less in the same range 
depending on Mn concentration 
\cite{Fukuma2003-1,Fukuma2008,Zvereva2010}.  T$_c$ values reported for RS or nearly RS Ge$_{1-x}$Cr$_x$Te films are higher\cite{Fukuma2006-2}. 
Note that the lattice parameters reported for $x=0.24$ and 0.55 in Ge$_{1-x}$Mn$_x$Te are 5.939 and
5.895 \AA, respectively, which compare well with  our calculated values 5.964 and 5.914 \AA, for $x=0.25$ and 0.50,  respectively (Table \ref{I}).

\begin{figure}
\center \includegraphics[width=7cm]{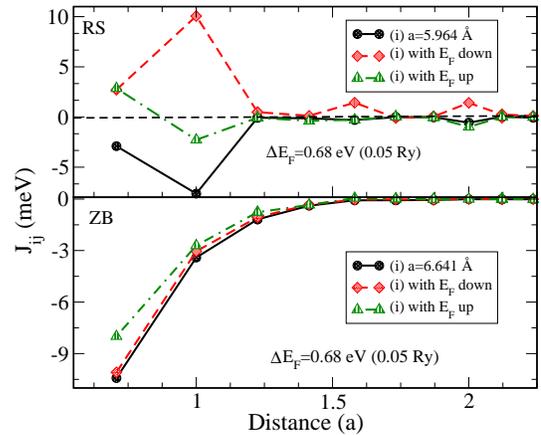}                     
\caption {(Color online) Change (non self-consistent result) in Mn-Mn exchange interactions  in RS and ZB Ge$_{0.75}$Mn$_{0.25}$Te 
computed by moving the Fermi energy (rigid shift) by 0.68 eV (0.05 Ry) up or down with respect to the result of the self-consistent
 calculation for the alloy. Solid lines with solid circles are the results of self-consistent calculations for the
equilibrium lattice parameters.
}
\label{f12}
\end{figure}

\begin{figure}
\center \includegraphics[width=7cm]{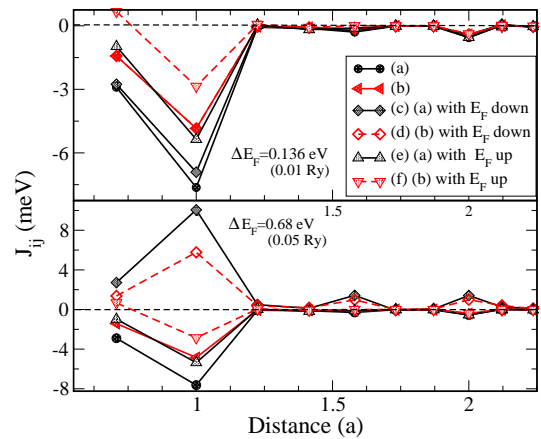}                        
\caption {(Color online) Change (non self-consistent result) in Mn-Mn exchange interactions  in RS Ge$_{0.75}$Mn$_{0.25}$Te 
computed by moving the Fermi energy (rigid shift) by 0.136 eV (0.01 Ry) and 0.68 eV (0.05 Ry) up or down with respect to the result of the self-consistent
 calculation for the alloy. Results are shown for two different lattice parameters: (a) equilibrium lattice parameter 5.964 \AA,
(b) expanded lattice parameter 6.125 \AA.
}
\label{f13}
\end{figure}

\begin{figure}
\center \includegraphics[width=7cm]{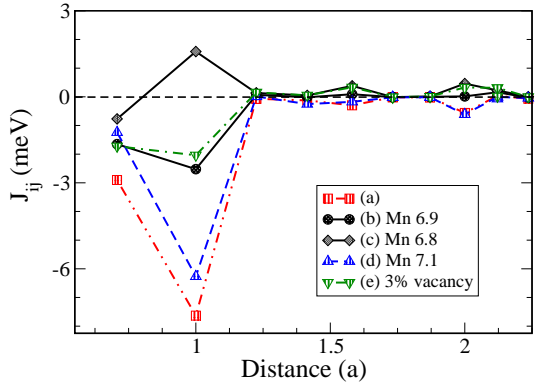}                        
\caption {(Color online) Change in Mn-Mn exchange interactions brought about by change in carrier concentration in various different ways in RS Ge$_{0.75}$Mn$_{0.25}$Te.
(a) Self-consistent result for equilibrium lattice parameter 5.964\AA, (b) same as (a), but by assigning 6.9 electrons to Mn
atoms, thus simulating hole-doping, (c) same as (b), but by assigning 6.8 electrons to Mn
atoms, thus simulating larger hole concentration, (d) same as (b) or (c), except that Mn valence has been moved to 7.1, simulating electron-doping, 
(e) self-consistent result for equilibrium lattice parameter, but by replacing 3\% of the atoms in the Ge-Mn sublattice
by empty spheres (vacancies).
}
\label{f13a}
\end{figure}

Our zero temperature results indicate that exactly at the
concentration $x$=0.25, Ge$_{1-x}$Mn$_x$Te should be either a semiconductor with small gap or at 
best a zero gap semiconductor, in both RS and ZB structures. 
The magnetic structure of the ground state in this case is not FM, it is either AFM or more complex. 
At finite temperatures, free carriers in the conduction band and holes
in the valence band are expected and the number of such carriers may be significant in view of the 
fact that the gap is close to zero. However, it is believed that
there are also a large number of carriers, mainly holes, with temperature-independent 
concentration\cite{Fukuma2003-1}. RS GeTe is reported to be a narrow gap semiconductor
with a band gap of about 0.2 eV\cite{Tsu}. Its very high $p$-type conduction is ascribed to large 
cation (Ge) vacancies\cite{Fukuma2008b,Tsu,Lewis,Damon}, which seem to
dominate over transport due to thermally excited carriers.
 We have thus decided to explore this issue, in the 
framework of our zero temperature formalism, in several ways. Guided by the fact that experiments 
indicate presence of uncompensated carriers, with 
temperature-independent concentrations, we first explore the effects of such carriers using 
the simplest possible approach. The easiest thing for us is to find the 
 change in Mn-Mn exchange interaction by moving the Fermi energy
up (simulating holes) or down (simulating electrons). This is a one step non-self-consistent 
(frozen potential) calculation, in the spirit of the rigid band model.
After self-consistency has been achieved for the equilibrium lattice parameter, exchange 
interactions are calculated for the self-consistent potential and the correct
Fermi energy, and then also
for the same potential but assuming the Fermi energy to be slightly higher/lower to simulate holes/electrons.
 The Fermi energy coming out from the self-consistent calculation
for the equilibrium lattice parameter is simply moved up or down by 0.136/0.68 eV (0.01/0.05 Ry). In Fig.\ref{f12} 
we show this change for both ZB and RS structure Ge$_{0.75}$Mn$_{0.25}$Te
and $\Delta(E_F)=\pm 0.68$ eV (0.05 Ry).
In both cases, i.e. Fermi energy changed by $\pm$0.68 eV, there is an increase in the Mn-Mn 
interaction.
In Fig.\ref{f13} we compare the results of changing he Fermi energy by 0.136 eV  and  0.68 eV  in 
RS Ge$_{0.75}$Mn$_{0.25}$Te for two different lattice parameters.  
This indicates that both electron- and hole-doping of the system would be
an efficient way to drive it toward ferromagnetism. 
 Of course, hole- and electron-doping can be simulated in other ways as well, for example, by changing the effective valence of
  the atoms. This can be done by changing the valence of the magnetic or the nonmagnetic atoms. 
In Fig.\ref{f13a} we show the effect of altering the valence of
Mn atoms from 7 to 6.8, 6.9 and 7.1 for different lattice parameters. This is equivalent to studying 
the effect of alloying of Mn with other elements to the left
or right in the same row of the periodic table using virtual crystal approximation (VCA). 
These calculations are self-consistent, as opposed to the above results with rigidly
shifted Fermi levels, but may suffer from the weaknesses of VCA. We have also included the case 
where Mn concentration in the Ge-sublattice is reduced from 25\% to 22\% and 
the remaining 3\% of the sites on this sublattice are left vacant. In all these calculations 
it is the magnetic atom Mn, whose concentration is effectively altered to create the
electrons or holes. Next, we explore the effect of creating holes or electrons by keeping the concentration of Mn atoms fixed at 25\%. 

\begin{figure}
\center \includegraphics[width=7cm]{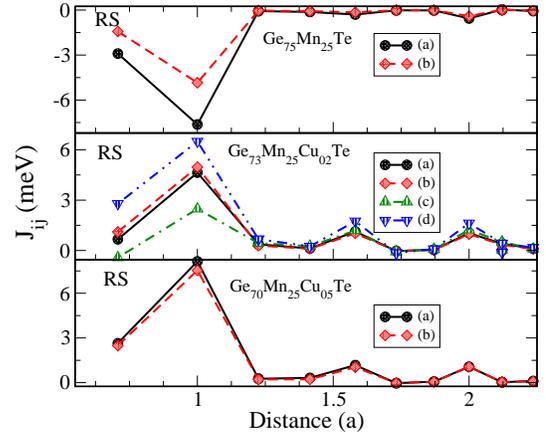}                   
\caption {(Color online) Mn-Mn exchange interactions in RS Ge-Mn-Te, obtained by putting 2\% and 5\% Cu in the 
Ge-Mn sublattice. Cu-atoms randomly replace the Ge-atoms, thus keeping the
Mn concentration fixed at 25\%. The legends (a)-(d) refer to different lattice parameters: 
(a) equilibrium Ge$_{0.75}$Mn$_{0.25}$Te lattice parameter 5.964 \AA, 
(b) expanded lattice parameter 6.125 \AA, (c) and (d) contracted lattice parameters 5.750 \AA$\;$ and 5.534 \AA, respectively.
}
\label{f14}
\end{figure}

We study separately  the effect 
of introducing holes by  doping the Ge-sublattice and the Te-sublattice. To create holes in the 
Ge-sublattice, we replace some of the Ge-atoms with Cu, and to create holes in the
Te-sublattice we replace some Te-atoms with Sn or simply vacancies. Results are shown in Figs. 
\ref{f14} and \ref{f15}. Fig. \ref{f14} shows the effect of replacing some of the
Ge atoms with Cu, thus creating holes. Results show the effect of increasing hole-concentration 
as well as the effect of changing the lattice parameter or the volume per atom.
We have considered only the RS case, as all the available experimental results are for this 
structure. It is clear that even a reasonably low level of Cu-doping of the 
Ge-sublattice can substantially increase the Mn-Mn exchange interaction, changing it from 
negative to positive and explaining the observed ferromagnetism of Ge$_{1-x}$Mn$_x$Te
thin films. Fig. \ref{f15} shows the effect of creating holes via doping the Te-sublattice with 
Sn or creating vacancies in this sublattice. Again, there is a change in the
Mn-Mn interaction in the direction of ferromagnetism (decreasing antiferromagnetic interactions).

\section{Summary of results and conclusions}

\begin{figure}
\center \includegraphics[width=7cm]{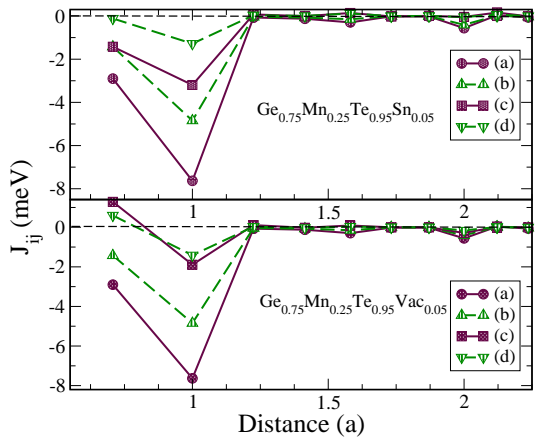}            
\caption {(Color online) Mn-Mn exchange interactions in RS Ge$_{0.75}$Mn$_{0.25}$Te$_x$Z$_{1-x}$, where Z stands for 
Sn-atoms or vacancies. Sn-atoms or vacancies occupy random positions 
solely in the Te-sublattice. Legends: (a) $x$=1.0, a= 5.964 \AA, (b) $x$=1.0, a=6.286 \AA, (c) $x$=0.95, a= 5.964 \AA, (d) $x$=0.95, a=6.286 \AA.
}
\label{f15}
\end{figure}

We have examined theoretically the possibility of half-metallic ferromagnetism in bulk Ge$_{1-x}$TM$_x$Te 
alloys, with TM being the transition metals V, Cr and Mn. 
FP-LAPW calculations reveal the possibility of half-metallicity for some of the ordered alloys. 
For these cases we have calculated the exchange interactions and the Curie
temperatures. The later calculations apply to cases where the Ge and the TM atoms occupy 
randomly the Ge-sublattice. The effect of this disorder is taken into account using the
CPA.  Ferromagnetic interactions are strongest in the Cr-doped ZB GeTe.
The Cr-doping case is most promising, as in both RS and ZB structures FM interactions 
dominate, and these remain FM with changes in the lattice parameter.
The next promising case is V-doping in the ZB structure, the RS counterpart showing strong 
antiferromagnetic (AFM) interactions, particularly at and around the
equilibrium lattice parameter. These AFM interactions weaken on both sides of the equilibrium 
lattice parameter value and become FM only at much higher volume.
Our calculations for TM=Mn and $x$=0.25 shows the substance to be AFM (for the ordered as well 
as disordered compounds, with the gap being narrower in the disordered case).
We show that this AFM behavior is linked to the substance being a narrow/zero gap 
semiconductor/semimetal at this Mn concentration. We further establish that the presence
of uncompensated carriers should drive the material toward ferromagnetism. The origin of these 
carriers could be vacancies, impurity atoms, and/or structural imperfections. 
In fact both Ge-Te and Ge$_{1-x}$TM$_x$Te samples are known to have large number of holes 
($p$-type carriers) due to cation(Ge) vacancies\cite{Fukuma2008b,Tsu,Lewis,Damon}.
A comparison of the results in Figs. \ref{f14} and \ref{f15} clearly shows that holes in 
the Ge-Mn sublattice is  much more effective in  driving the system to ferromagnetism
than the holes in Te-sublattice. This seems to be consistent with  the experimental 
observation that all ferromagnetic films of Ge$_{1-x}$TM$_x$Te have a large number of temperature-independent 
p-type carriers, associated with vacancies in the Ge sublattice.
The calculated
exchange interactions due to carriers originating from these causes are sufficient to 
account for the observed ferromagnetism in thin films of Ge$_{1-x}$Mn$_x$Te.
Our results for the exchange interactions are consistent with experiments in that the highest 
T$_c$ reported so far seem to be for thin films of Ge$_{1-x}$Cr$_x$Te.
Some differences between our calculated results and the experiments could be ascribed to the 
fact that our calculations are for bulk alloys, while experimental results
are mostly for thin films. For example, distortions with respect to the bulk structures 
may be present in thin films, in addition 
to the fact that the lattice parameters of the thin films are usually larger than their bulk counterparts.\cite{Fukuma2008}

ACKNOWLEDGMENTS

The work of Y.L. and S.K.B. was supported by a grant from the Natural Sciences and Engineering Research Council of Canada.
Computational facilities for this work was provided by SHARCNET, Canada. The work of Y.L. was also partly supported by the Natural Science
Foundation of China (No.$10974228$). The work of J.K. was supported by a grant from the Czech Science Foundation (202/09/0775).


\begin{thebibliography}{99}
\bibitem{Fukuma2001} Y. Fukuma, T. Murakami, H. Asada, and T. Koyanagi, Physica E {\bf 10}, 273 (2001).
\bibitem{Fukuma2003-1} Y. Fukuma, H. Asada,, J. Miyashita, N. Nishimura, and T. Koyanagi, J. Appl. Phys. {\bf 93}, 7667 (2003).
\bibitem{Fukuma2003-2} Y. Fukuma, N. Nishimura, F. Odawara, H. Asada, and T. Koyanagi, J. Supercond: Incorp. Nov. Magn.,
 {\bf 16}, 71 (2003).
\bibitem{Fukuma2006-1} Y. Fukuma, T. Taya, S. Miyawaki, T. Irisa, H. Asada, and T. Koyanagi, J. Appl. Phys. {\bf 99}, 08D508 (2006).
\bibitem{Fukuma2006-2} Y. Fukuma, H. Asada, T. Taya, T. Irisa, and T. Koyanagi, Appl. Phys. Lett. {\bf 89}, 152506 (2006).
\bibitem{Fukuma2006-3} Y. Fukuma, H. Asada, and T. Koyanagi, Appl. Phys. Lett. {\bf 88}, 032507 (2006).
\bibitem{Fukuma2008} Y. Fukuma, H. Asada, S. Miyawaki, T. Koyanagi, S. Senba, K. Goto, and H. Sato, Appl. Phys. Lett.{\bf 93},
252502 (2008).
\bibitem{Chen2008} W.Q. Chen, S.T. Lim, C.H. Sim, J.F. Bi, K.L. Teo, T. Liew, and T.C. Chong, J. Appl. Phys. {\bf 104}, 063912 (2008).
\bibitem{Zhao2006} Y-H Zhao, W-H Xie, L-F Zhu, and B-G Liu, J. Phys.: Condens. Matter {\bf 18}, 10259 (2006).
\bibitem{Ciucivara2007} A. Ciucivara, B.R. Sahu, and L. Kleinman, \prb {\bf 75}, 241201(R) (2007).
\bibitem{Fukuma2007} Y. Fukuma, H. Asada, N. Moritake, T. Irisa, and T. Koyanagi, Appl. Phys. Lett. {\bf 91}, 092501 (͑2007͒).
\bibitem{Lim2009} S. T. Lim, J. F. Bi, K. L. Teo, Feng Y. P, T. Liew,and T. C. Chong, Appl. Phys. Lett. {\bf 95}, 072510 (͑2009͒).
\bibitem{Lechner2010} R. T. Lechner \etal, Appl. Phys. Lett. {\bf 97}, 023101 (͑2010͒).
\bibitem{Tong2011} F. Tong, J. H. Hao, Z. P. Chen, G. Y. Gao, and X. S. Miao, Appl. Phys. Lett. {\bf99}, 081908 (2011).
\bibitem{Fukuma2008b} Y. Fukuma \etal, J. Appl. Phys. {\bf 103}, 053904 (͑2008͒).
\bibitem{Zvereva2010} E. A. Zvereva \etal, J. Appl. Phys. {\bf 108}, 093923 (͑2010͒).
\bibitem{Lim2011a} S. T. Lim, J. F. Bi, K. L. Teo, and T. Liew, J. Appl. Phys. {\bf 109}, 07C314 (2011).
\bibitem{Lim2011b} S. T. Lim, J. F. Bi, Lu Hui, and K. L. Teo, J. Appl. Phys. {\bf 110}, 023905 (2011).
\bibitem{Lim2011c} S. T. Lim, L. Hui, J. F. Bi, and K. L. Teo, J. Appl. Phys. {\bf 110}, 113916 (2011).
\bibitem{Hassan2011} M. Hassan \etal, J. Cryst. Growth. {\bf 323}, 363 (2011).
\bibitem{Kilanski2010} L. Kilanski \etal, \prb {\bf 82},094427(͑2010͒).
\bibitem{Sato2} K. Sato \etal, Rev. Mod. Phys. {\bf 82}, 1633 (2010).
\bibitem{Bergqvist} L. Bergqvist, O. Eriksson, J. Kudrnovsk\'{y}, V. Drchal, P. Korzhavyi, and I. Turek, \prl {\bf 93}, 137202 (2004).
\bibitem{Blaha-wien2k} P. Blaha, K. Schwarz, G.
Madsen, D. Kvasnicka, J. Luitz, WIEN2k, An Augmented Plane Wave Plus
Local Orbitals Program for Calculating Crystal Properties, Vienna
University of Technology Inst. of Physical and Theoretical, Vienna,
Austria, 2008; P. Blaha, K. Schwarz, P. Sorantin, Comput.
Phys.Commun. {\bf 59}, 399 (1990).
\bibitem{Kudrnovsky1990} J. Kudrnovsk\'y and V. Drchal, \prb {\bf 41}, 7515 (1990).
\bibitem{Turek97} I. Turek, V. Drchal, J. Kudrnovsk\'y, M. \v{S}ob, and
P. Weinberger, {\it Electronic Structure of Disordered Alloys,
Surfaces and Interfaces} (Kluwer, Boston-London-Dordrecht, 1997).
\bibitem{Kohn-pr-136-B864-1964}  P. Hohenberg and W. Kohn, Phys. Rev.  {\bf 136},
B864 (1964); W. Kohn and L.J. Sham, Phys. Rev. {\bf 140}, A1133
(1965).
\bibitem{Perdew-prb-45-13244-1992} J.P. Perdew, Y. Wang, Phys. Rev. B {\bf 45}, 13244 (1992).
\bibitem{perdew-prl-77-3865-1996}  J.P. Perdew, K. Burke and M. Ernzerhof,
Phys. Rev. Lett. {\bf 77}, 3865 (1996).
\bibitem{Monkhorst-prb-13-5188-1976} H.J. Monkhorst, J.D. Park, Phys. Rev. B {\bf 13},
5188 (1976).
\bibitem{Murnaghan-pnas-30-5390-1944} F.D. Murnaghan, Proc. Natl. Acad. Sci. USA {\bf 30},
5390 (1944).
\bibitem{VWN} S.H.Vosko, L. Wilk, and M. Nusair, Can. J. Phys. {\bf 58}, 1200 (1980).
\bibitem{Sandratskii} L.M. Sandratskii, R. Singer, and E. Sasio\u{g}lu, \prb {\bf 76}, 184406 (2007).
\bibitem{Pajda2001} M. Pajda, J. Kudrnovsk\'y, I. Turek, V. Drchal,
and P. Bruno, Phys. Rev. B {\bf 64}, 174402 (2001).
\bibitem{Bose2010} S.K. Bose and J. Kudrnovsk\'y, \prb {\bf 81}, 054446 (2010).
\bibitem{Liu2010}Y. Liu, S.K. Bose, J. Kudrnovsk\'y, \prb {\bf 82}, 094435 (2010).
\bibitem{Liechtenstein84I} A.I. Liechtenstein, M.I. Katsnelson and
V.A. Gubanov, J. Phys.F: Met.Phys. {\bf14}, L125 (1984).
\bibitem{Liechtenstein84II} A.I. Liechtenstein, M.I. Katsnelson and
V.A. Gubanov, Solid. State. Commun. {\bf 54}, 327 (1985).
\bibitem{Gubanov92} V.A. Gubanov, A.I. Liechtenstein, A.V. Postnikov
{\em Magnetism and the electronic structure of crystals}, edited by
M. Cardona, P. Fulde, K. von Klitzing, H.-J. Queisser (Springer, Berlin, 1992).
\bibitem{Turek2006} I. Turek, J. Kudrnovsk\'y, V. Drchal,  and P. Bruno,
Philos. Mag. {\bf 86}, 1713 (2006).
\bibitem{Prange} C.S. Wang, R.E. Prange, and V. Korenman,  Phys. Rev. B {\bf 25}, 5766 (1982).
\bibitem{Heine1} V. Heine, J.H. Samson, and C.M.M. Nex, J. Phys. F: Met. Phys. {\bf 11}, 2645 (1981).
\bibitem{Heine2} V. Heine and J.H. Samson, J. Phys. F: Met. Phys. {\bf 13}, 2155 (1983).
\bibitem{Tsu} R. Tsu, W.E. Howard, and L. Esaki, Phys. Rev. {\bf 172}, 779 (1968).
\bibitem{Lewis} J.E. Lewis, Phys. Status Solidi B {\bf 59}, 367 (1973).
\bibitem{Damon} D.H. Damon, M.S. Lubell, and R. Mazelsky, J. Phys. Chem. Solids {\bf  28}, 520 (1967).
\end {thebibliography}
\end{document}